\documentclass[twocolumn]{webofc}

\usepackage[varg]{txfonts}   
\usepackage{hyperref}
\usepackage{url}
\hypersetup{colorlinks=true,citecolor=blue,urlcolor=blue,linkcolor=blue}
\newcommand\beq{ \begin{eqnarray} }
\newcommand\eeq{ \end{eqnarray} }

\begin{document}
\title{Speed of sound exceeding the conformal bound in dense QCD-like theories}
%
%

\author{\firstname{Etsuko} \lastname{Itou}\inst{1,2}\fnsep\thanks{\email{itou@yukawa.kyoto-u.ac.jp}} \and
        \firstname{Kei} \lastname{Iida}\inst{3,4} \and
        \firstname{Kotaro} \lastname{Murakami}\inst{2,5} \and
        \firstname{Daiki} \lastname{Suenaga}\inst{6}
}

\institute{Yukawa Institute for Theoretical Physics, Kyoto University, 
Kitashirakawa Oiwakecho, Sakyo-ku, Kyoto 606-8502 Japan
\and
RIKEN Center for Interdisciplinary Theoretical and Mathematical Sciences (iTHEMS), RIKEN,
2-1 Hirosawa, Wako, Saitama 351-0198 Japan 
\and
Department of Liberal Arts, The Open University of Japan, Chiba 261-8586, Japan
\and
RIKEN Nishina Center, RIKEN, 2-1 Hirosawa, Wako, Saitama 351-0198, Japan
\and
Department of Physics, Institute of Science Tokyo, 2-12-1 Ookayama, Megro, Tokyo 152-8551, Japan 
\and
Kobayashi-Maskawa Institute for the Origin of Particles and the Universe, Nagoya University, Nagoya, 464-8602, Japan
          }

\abstract
{We investigated the phase structure and the equation of state (EoS) for dense two-color QCD at low temperatures using the lattice Monte Carlo simulations. A rich phase structure below the pseudo-critical temperature 
$T_c$ as a function of quark chemical potential 
 has been revealed. In a high-density regime, we can see a superfluid phase, where the diquark condensate takes a non-zero expectation value. We have newly found that the speed of sound exceeds the conformal bound, which is the value of the relativistic free theory.
This talk is based on Refs.~\cite{Iida:2022hyy, Iida:2024irv, Itou:2025vcy}.
}
\maketitle
%
\section{Introduction}
\label{intro}
Protons and neutrons are described by quantum chromodynamics (QCD), a non-abelian gauge theory based on the SU(3) gauge symmetry that governs the strong interaction. Among the available formulations, lattice gauge theory provides the only known regularization that is both nonperturbative and manifestly gauge invariant. Numerical simulations performed by incorporating the QCD action into lattice Monte Carlo methods 
have achieved precise agreement with experimental measurements. As a result, lattice Monte Carlo simulation is widely recognized as an ab initio approach to the study of general gauge theories. It serves as a controlled numerical experiment, enabling investigations in regimes where analytic methods are not applicable.

The equation of state (EoS) of QCD at zero quark-chemical-potential ($\mu$) has been extensively explored through such first-principles lattice calculations. The temperature dependence of thermodynamic quantities in this region, corresponding to the $\mu=0$ axis of the QCD phase diagram, is now well established. Both the pressure ($p$) and the energy density ($e$) increase monotonically with temperature after the chiral crossover. The corresponding sound velocity ($c_{\mathrm s}$) exhibits a characteristic minimum around the pseudo-critical temperature ($T_c$) and then increases monotonically at higher temperatures.

Here, we would like to focus on a less understood part of the phase diagram by examining 
the $\mu$-dependence of EoS and sound velocity at low temperature. 
In both the low-density and asymptotically high-density regimes, various model analyses provide qualitative guidance. However, the most important region for understanding dense QCD, namely the intermediate-density regime, remains theoretically uncertain, and lattice calculations are strongly desired. The corresponding sound velocity is expected to rise from its low-density behavior and eventually approach the relativistic limit $c_{\mathrm s}^2/c^2 = 1/3$ at extremely high density.

A major obstacle in exploring this dense regime is the sign problem: the fermion determinant becomes complex at finite chemical potential, rendering the Monte Carlo approach impractical. The sign problem is known to be NP-hard, meaning that no efficient solution is expected within the standard Monte Carlo framework~\cite{Nagata:2021ugx}. To make progress, one must either modify the theory, for example by studying QCD-like theories without a sign problem, or seek new computational paradigms such as quantum algorithms.
Here, we follow the former strategy: we consider two-color QCD, which is a QCD-like theory. Owing to the pseudoreality of the fermion representation, this theory does not suffer from the sign problem even at finite density, allowing one to apply standard Monte Carlo methods.

On the other hand, in recent years, an intriguing prediction has emerged from phenomenological analyses and effective models of dense QCD matter~\cite{Masuda:2012ed, McLerran:2018hbz, Kojo:2021ugu,Kawaguchi:2024iaw}: the possibility that the sound velocity develops a pronounced peak at intermediate baryon densities. Studies motivated by observations of neutron star masses (and radii) suggest that a smooth quark–hadron crossover, consistent with the astrophysical data, may lead to a maximum of $c_{\mathrm s}^2/c^2$. Similar behavior has also been proposed within the quarkyonic matter framework and 
other effective models of dense QCD.
Furthermore,  it has also been proposed that the peak may signal the onset of quark saturation in dense matter and this mechanism could operate for any number of colors~\cite{Kojo:2021hqh}. It is based on some effective model analysis, but it is worth studying the lattice calculation of the two-color QCD case.

To summarize the situation, two-color QCD shares many essential nonperturbative features with real QCD, such as confinement, chiral symmetry breaking, and characteristic topological structures at $\mu = 0$ where lattice Monte Carlo simulations are fully reliable. 
At finite density, effective-model analyses suggest that similar qualitative behavior may also persist. However, direct first-principles studies of three-color QCD at finite density are severely limited by the sign problem. In this situation, lattice simulations of two-color QCD at finite density offer one of the few reliable, first-principles approaches to the dynamics of dense QCD, and therefore play an essential role in clarifying which features of dense matter survive beyond effective-model descriptions.


\section{Two-color QCD phase diagram}
\label{sec:phase-diagram}
\begin{figure}[h]
\centering
\includegraphics[width=8cm,clip]{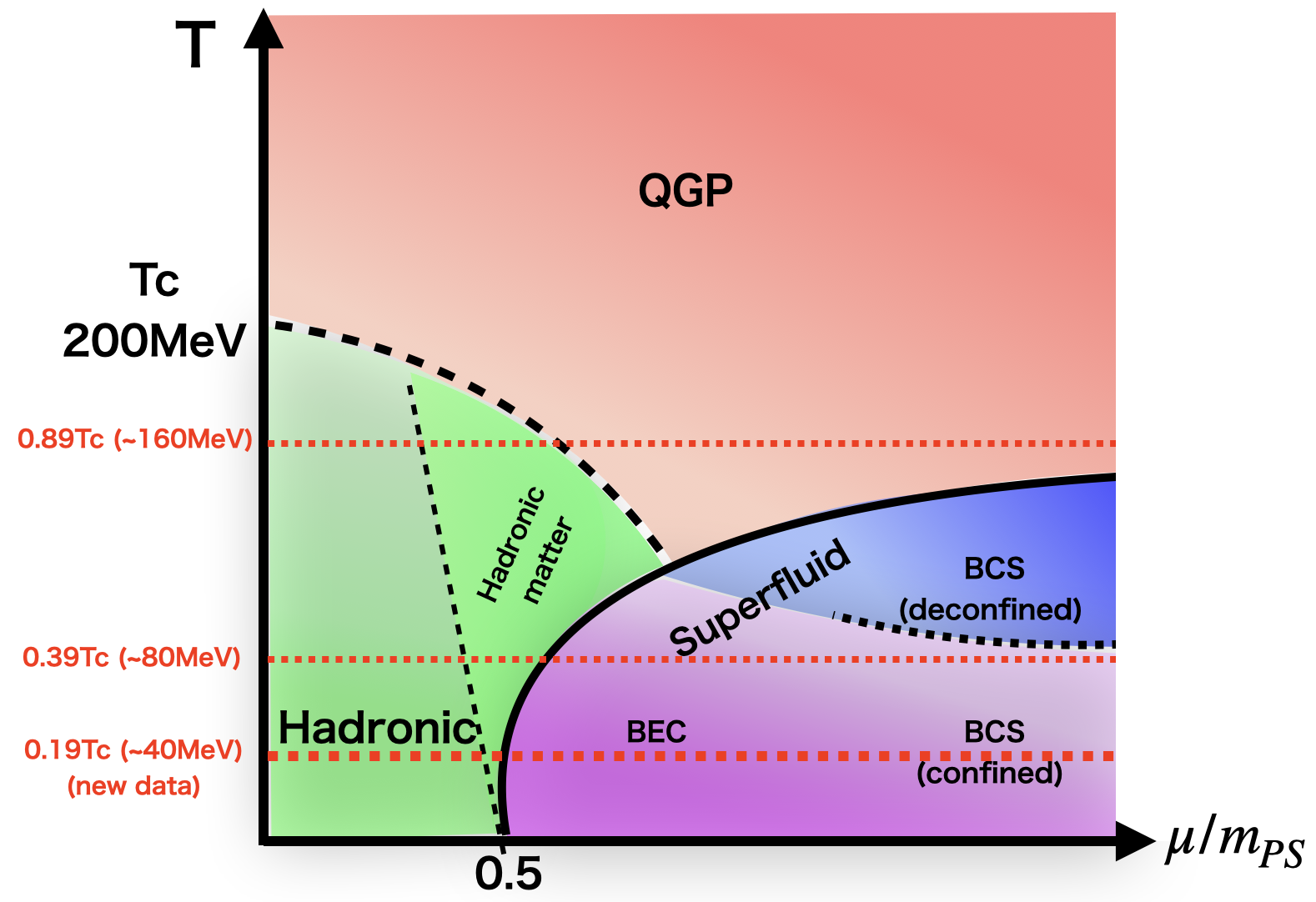}
\caption{Two-color QCD phase diagram. The data at $160$ MeV and $80$ MeV are originally presented in Refs.~\cite{Iida:2019rah, Iida:2022hyy}, while the ones at $40$ MeV are obtained in Ref.~\cite{Iida:2024irv}.}
\label{fig:phase}     
\end{figure}

Before presenting the results for the EoS, let us briefly introduce the current understanding of the two-color QCD phase diagram. In recent years, at least four independent lattice groups have been actively studying this theory at finite density (see Ref.~\cite{Itou:2025vcy} and references therein), and their findings can be summarized in Fig.~\ref{fig:phase}. One notable point is that the onset of the superfluid phase occurs at temperatures as high as approximately 100 MeV, which is significantly higher than what had been anticipated in earlier studies.

\begin{figure}[h]
\centering
\includegraphics[width=10cm,clip]{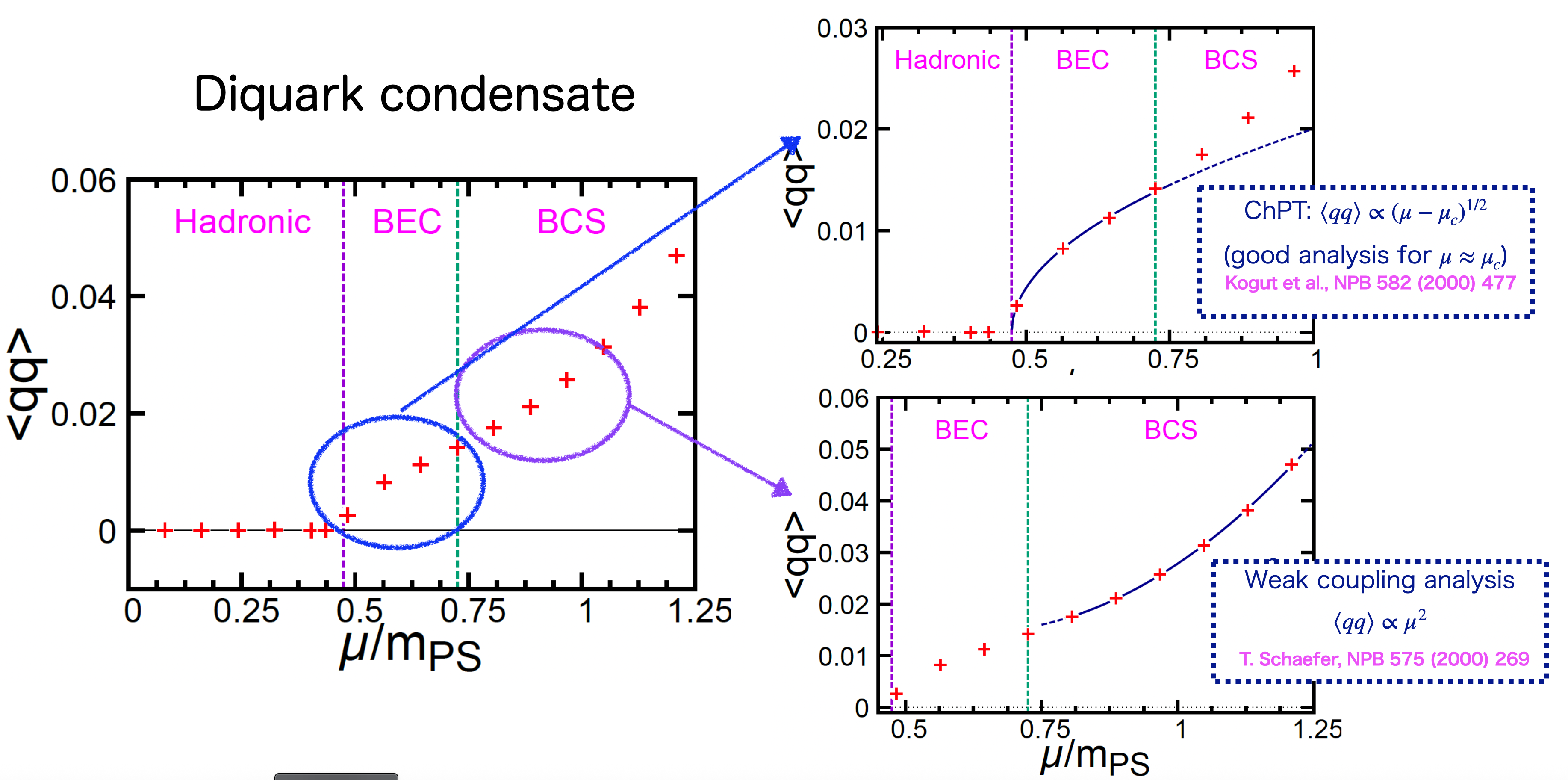}
\caption{Our lattice data of the diquark condensate as a function of $\mu$. The data were originally obtained in Ref.~\cite{Iida:2024irv}.}
\label{fig:diquark}     
\end{figure}

In more detail, focusing especially on the low-temperature region, the transition from the hadronic (normal) phase to the superfluid phase takes place around the chemical potential corresponding to half the mass of the lightest hadron excitation, namely the pseudo-scalar meson ($m_{\rm PS}$).  Beyond this point, $\mu_c \approx m_{\rm PS}/2$, the diquark condensate, $\langle qq \rangle$, takes nonzero expectation values, which is indeed the order parameter of superfluidity, as shown in Fig.~\ref{fig:diquark}.
Around $\mu_c$, the quark number density also begins to rise from zero, and the system undergoes spontaneous breaking of the baryon number symmetry. This is the characteristic feature of the diquark superfluid phase in two-color QCD.

Furthermore, in the superfluid region, the system exhibits a smooth Bose–Einstein condensate(BEC)–BCS crossover. Here, we determine a guiding line (green-dotted vertical line) of BEC-BCS crossover by seeing the quark number density. Thus, if the lattice numerical data of the quark number density can be well-approximated by the tree-level lattice calculation, then we identify such a region as the BCS phase.
At lower chemical potentials in the superfluid phase, the diquark condensate behaves as a BEC of tightly bound hadronic states. In contrast, at higher densities it gradually evolves into a BCS-type pairing of weakly bound quarks because of the asymptotic freedom of QCD-like theories. 
The scaling behavior of the diquark condensate (see the right panels in Fig.~\ref{fig:diquark}) provides a clear signal for this crossover, and our lattice data are consistent with the prediction of mean-field chiral perturbation theory (ChPT) in the BEC phase~\cite{Kogut:2000ek}, while the data can be fitted by a quadratic function of $\mu$ in the BCS phase, which is expected by the weak coupling analysis~\cite{Schafer:1999fe, Hanada:2011ju, Kanazawa:2013crb}.

\section{Equation of state and sound velocity}
\label{sec:eos}
Let us now turn to the EoS of two-color QCD at finite density. As a first step, it is helpful to examine the low-density region, where the system is in the BEC phase. In this regime, ChPT provides an effective description, and indeed, our lattice data for the diquark and chiral condensates in the BEC region are consistent with these predictions.
The thermodynamic quantities have been obtained by these analytic studies~\cite{Son_2001, Hands:2006ve};
\beq
p_{\rm ChPT}&=&4N_f F^2 \mu^2 \left( 1- \frac{\mu_c^2}{\mu^2} \right)^2,  \nonumber\\
e_{\rm ChPT}&=&4N_f F^2 \mu^2 \left( 1- \frac{\mu_c^2}{\mu^2} \right) \left( 1+3 \frac{\mu_c^2}{\mu^2} \right). \label{eq:ChPT-e} 
\eeq
Here, $F$ denotes the pion decay constant ($F=f_\pi/2$) in two-color QCD.

\begin{figure}[h]
\centering
\includegraphics[width=5cm,clip]{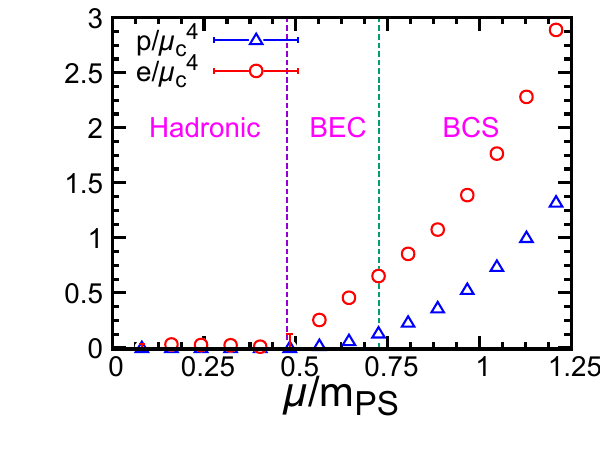}\\
\includegraphics[width=5cm,clip]{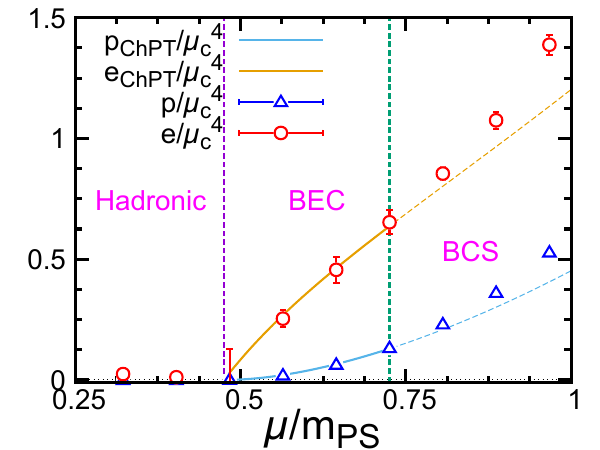}
\caption{Our data for EoS as a function of $\mu$. The data were originally obtained in Ref.~\cite{Iida:2024irv}.}
\label{fig:eos}       
\end{figure}
As shown in Fig.~\ref{fig:eos}, our data can be well-fitted by these functional forms, and the best fit values were obtained as $F= 51.1(5)$ MeV and $F=56.7(7)$ MeV from the fits of $p/\mu_c^4$ and $e/\mu_c^4$, respectively.
These values are similar to the earlier result, $F=60.8(1.6)$ MeV, obtained by independent simulation using the rooted staggered fermions in Ref.~\cite{Astrakhantsev:2020tdl}.

\begin{figure}[h]
\centering
\includegraphics[width=8cm,clip]{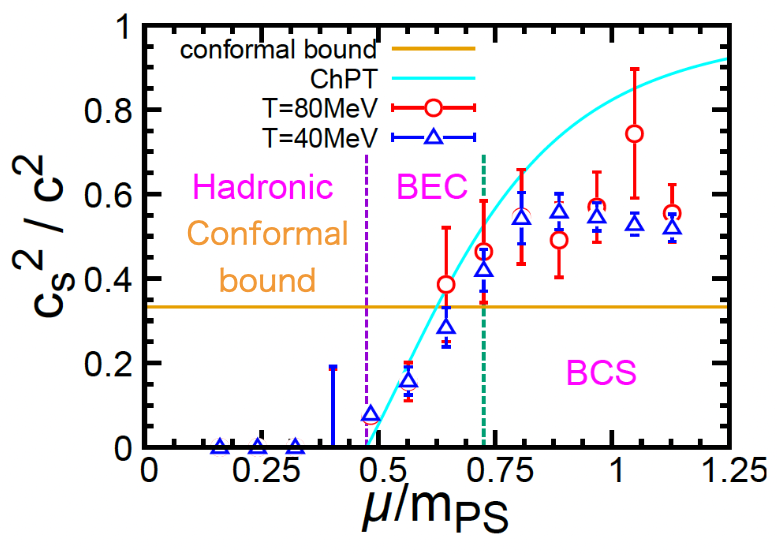}
\caption{Sound velocity squared as a function of $\mu$. The plot is originally shown in Ref.~\cite{Iida:2024irv}.}
\label{fig:sound-velocity}     
\end{figure}

Finally, let us show our results for the sound velocity.
Figure~\ref{fig:sound-velocity} depicts the results 
at $T=40$ MeV and $80$ MeV. It is evaluated by 
\beq
c_\mathrm{s}^2 (\mu)/c^2 &=& \Delta p (\mu)/\Delta e (\mu)\nonumber\\
&=& \frac{ p(\mu +\Delta \mu) - p(\mu -\Delta \mu)}{e(\mu +\Delta \mu) - e(\mu -\Delta \mu)}
\eeq
An important observation is that the temperature dependence of the sound velocity is essentially negligible within the range we have studied. This indicates that, in two-color QCD, the dominant density dependence of the EoS determines the behavior of the sound velocity, while thermal effects play only a minor role; the usual isentropic sound velocity would be very close to the isothermal one evaluated here.

In the BEC phase, the calculated sound velocity is fully consistent with the predictions of ChPT provides,
\beq
c_{\mathrm s}^2/c^2=(1-\mu_c^4/\mu^4)/(1+3\mu_c^4/\mu^4),\label{eq:cs-ChPT}
\eeq
which is a parameter-free expression once we fix the value of $\mu_c$.
Our lattice data follow this behavior remarkably well.

At higher densities, however, our data show a discrepancy from the ChPT prediction, which approaches $1$ in the high-density limit. This deviation suggests that the ChPT analysis is no longer applicable in the BCS phase.
On the other hand, our data exhibit a striking trend: $c_{\mathrm s}^2/c^2$ exceeds the conformal bound, namely  $1/3$. 
This enhancement indicates the growing stiffness of the EoS as the system moves toward the BCS side of the crossover. 
Furthermore, some lattice Monte Carlo studies for three-color QCD with isospin chemical potential~\cite{Brandt:2022hwy, Abbott:2024vhj} and another study for two-color QCD~\cite{Hands:2024nkx} also reported similar results recently.
These results provide a controlled first-principles indication that dense QCD-like theories can realize a sound velocity larger than the conformal value, offering a new insight into the qualitative behavior expected in real QCD at intermediate densities.

\section{Summary}
\label{sec:Summary}
Lattice simulations of QCD-like theories that do not suffer from the sign problem have continued to advance our understanding of dense QCD matter. Two-color QCD at finite density and three-color QCD with nonzero isospin chemical potential can be treated with exact Monte Carlo algorithms, allowing us to investigate their thermodynamics from first principles. One of the central findings of the present study is that the sound velocity in two-color QCD can exceed the conformal bound $c_{\mathrm s}^2/c^2 = 1/3$. This result is noteworthy because, for roughly forty years, lattice Monte Carlo simulations of QCD at vanishing chemical potential had consistently satisfied $c_{\mathrm s}^2/c^2 \leq 1/3$ before our first study in Ref.~\cite{Iida:2022hyy}. These data therefore indicate that dense QCD-like theories may naturally accommodate a significantly stiffer equation of state, opening up a new possibility for the properties of dense QCD matter.

Possible implications for neutron star physics must, of course, be considered with appropriate caution. Nevertheless, if a large sound velocity is realized {\it even} in real QCD at intermediate densities, then the corresponding EoS could be considerably stiffer than conventionally assumed. Identifying the microscopic mechanism responsible for such behavior is an important challenge in quantum field theory. Several theoretical scenarios have been proposed (see Ref.~\cite{Itou:2025vcy} and references therein).

At the same time, progress is emerging from the effective theory community. By incorporating lattice constraints, more reliable models of dense QCD are being developed, offering a complementary avenue for exploring the intermediate-density regime that remains inaccessible to direct first-principles calculations~\cite{Fujimoto:2024pcd, Fukushima:2024gmp, Pasqualotto:2025kpo, Lopes:2025rvn}. The combination of lattice simulations, effective model building, and astrophysical observations will continue to play a crucial role in clarifying the structure of dense QCD matter.

%
%
%

\section*{Acknowledgments}
K.~I. and E.~I are supported by JSPS KAKENHI with Grant Number 25K01001. 
The work of E.~I. is supported by 
JSPS Grant-in-Aid for Transformative Research Areas (A) JP21H05190, 
JST SQAI with Grant Number JPMJPF2221,  
JST CREST with Grant Number JPMJCR24I3,  
and also supported by Program for Promoting Researches on the Supercomputer ``Fugaku'' (Simulation for basic science: from fundamental laws of particles to creation of nuclei) and (Simulation for basic science: approaching the new quantum era), and Joint Institute for Computational Fundamental Science (JICFuS), Grant Number JPMXP1020230411. 
This work is supported by Center for Gravitational Physics and Quantum Information (CGPQI) at Yukawa Institute for Theoretical Physics.
E.~I and D.~S are supported by JSPS KAKENHI with Grant Number 23H05439. 
D.~S. is also supported by JSPS KAKENHI No.~23K03377 and No.~25K17386.
K.~M. is supported in part by Grants-in-Aid for JSPS Fellows (Nos.\ JP22J14889, JP22KJ1870) and by JSPS KAKENHI with Grant No.\ 22H04917. 

\bibliography{2color}
%
%

\end{document}